%% file: eprint.tex
\documentclass[10pt, paper=a4, UKenglish]{article}
\usepackage{graphicx}
\def\Title#1{\begin{center} {\Large #1 } \end{center}}
\def\Author#1{\begin{center}{ \sc #1} \end{center}}
\def\Address#1{\begin{center}{ \it #1} \end{center}}

\newcommand\pubblock{\rightline{\begin{tabular}{l} Proceedings of the CTD/WIT 2019\\ \pubnumber\\
         \pubdate  \end{tabular}}}

\newenvironment{Abstract}{\begin{quotation} \begin{center} 
             \large ABSTRACT \end{center}\bigskip 
      \begin{center}\begin{large}}{\end{large}\end{center} \end{quotation}}

\newenvironment{Presented}{\begin{quotation} \begin{center} 
             PRESENTED AT\end{center}\bigskip 
      \begin{center}\begin{large}}{\end{large}\end{center} \end{quotation}}

\def\Acknowledgements{\bigskip  \bigskip \begin{center} \begin{large}
      \bf ACKNOWLEDGEMENTS \end{large}\end{center}}

\input econfmacros.tex

\usepackage{color}
\usepackage{lineno}
\usepackage{subfig}
\usepackage{hyperref}
\usepackage{amssymb}
\usepackage{tikz}
\usepackage{float}
\usepackage{multirow}
\usepackage{array}
\usepackage{colortbl}
\usepackage{hhline}
\usepackage[autostyle=true, style=british]{csquotes}

\newcommand\pubnumber{PROC-CTD19-090}

\newcommand\pubdate{\today}

\def\affiliation{
Institute for Astro- and Particle Physics \\
University of Innsbruck, Austria}

\newcommand{\conference}{Connecting the Dots and Workshop on Intelligent Trackers (CTD/WIT 2019)\\
Instituto de F\'isica Corpuscular (IFIC), Valencia, Spain\\ 
April 2-5, 2019}

\usepackage{fancyhdr}
\definecolor{mygrey}{RGB}{105,105,105}
\begin{document}

\large
\begin{titlepage}
\pubblock

\vfill
\Title{Track Seed Classification \\ with Deep Neural Networks}
\vfill

\Author{Felix Dietrich}
\Address{\affiliation}
\vfill

\begin{Abstract}
Future upgrades to the LHC will pose considerable challenges for traditional particle track reconstruction methods. We investigate how artificial Neural Networks and Deep Learning could be used to complement existing algorithms to increase performance. Generating seeds of detector hits is an important phase during the beginning of track reconstruction and improving the current heuristics of seed generation seems like a feasible task. We find that given sufficient training data, a comparatively compact, standard feed-forward neural network can be trained to classify seeds with great accuracy and at high speeds. Thanks to immense parallelization benefits, it might even be worthwhile to completely replace the seed generation process with the Neural Network instead of just improving the seed quality of existing generators.
\end{Abstract}

\vfill

\begin{Presented}
\conference
\end{Presented}
\vfill
\end{titlepage}
\def\thefootnote{\fnsymbol{footnote}}
\setcounter{footnote}{0}
%

\normalsize 


\section{Introduction}
\label{intro}

In the next decade, high-luminosity upgrades to the LHC will confront detectors with an order of magnitude increase in particle collisions. This will push the traditional track reconstruction software and hardware beyond their capabilities. Henceforth, upgraded or entirely new algorithms will be required to deal with the increased amount of data. The current track reconstruction approaches based on track seeding and track following allow for large contingency and hence are not optimal in terms of computational efficiency. Early fake classifications, especially during the first stages of track reconstruction, offer viable opportunities for a faster, more efficient reconstruction. Deep Neural Networks (DNNs) and Machine Learning provide an interesting and highly viable opportunity to combine multiple advantages into a single approach when it comes to the classification of such track seeds. With extremely high execution speed - thanks to inherent parallelizability - and a good chance to improve rejection rates for improper seeds due to the contingencies and heuristics of current algorithms, a DNN certainly seems up to the task. This approach is also underpinned by the surge of high performance and free to use deep learning frameworks, which have matured over the last years.

\section{Track Seeding}
\label{Track Seeding}
In the first stages of track reconstruction, a set of hits from the detector is basically sampled at random. This set is called a seed and starting from it, additional hits are found progressively while keeping the overall likelihood for a correct track in mind. This can, for example, be realized with a Kalman filter. However, since the space of possible seeds suffers from combinatorial explosion, it is necessary to constrain seeds by preselecting only those which are likely to result in a successful track candidate. To prevent falling into the combinatorial pitfall again, this pre-selection must be extremely performant and thus allows for little computational complexity. One way of doing it is by selecting three hits in the xy-plane perpendicular to the beam axis and then directly calculate the circle which they define. Since charged particles follow helical trajectories in the detector's magnetic field, they should lie on a circle in this plane. The distance between the circle and a possible primary vertex on the beam axis can then be used as a cut parameter to reduce the amount of possible seeds. It should still be noted that tracks with loosely constrained primary vertices might necessitate large cuts or even unconstrained seed generation, drastically increasing the number of eventual track candidates. But for many event types, an efficient initial seed generation could free up time for such computationally intensive searches. 

\ifx false
\section{Artificial Neural Networks}
\label{ANNs}
The first practical mathematical model intended to reproduce the way neurons work in the human brain has been around since at least the 1940s \cite{mcculloch}. The basic approach tries to mimic the structure and mechanisms observed in biology, effectively modelling the electro chemical interaction between neural cells. The most important observation is that a neuron receives signal inputs from many other neurons and ,once a certain threshold is reached, transmits the signal further down the network. This structure can be modelled mathematically using a directed, weighted graph. The human brain also contains much more complicated neural circuits which are subject of current research, but the simple directed graph model is already sufficient for an amazingly rich set of tasks. The perceptron, introduced in 1957 \cite{rosenblatt}, still forms the basis for most artificial neural network (ANN) models today:

\begin{equation}
\textbf{y}(\textbf{x}) = f(W \cdot \textbf{x})
\end{equation} 

where $\textbf{x}$ is an input vector, $f$ a non-linear function called activation function and $W$ a matrix containing the tunable weights of the model. While it may seem simple, it was proven \cite{approximationtheorem} that this model is already sufficient to approximate \emph{any} continuous function or discriminatory hypersurface on compact subsets of $\mathbb{R}^n$ with arbitrary accuracy. Training usually works by providing a set of inputs and comparing the model output to a known, true output. In that case the algorithm hence belongs to the class of supervised learning models. But it is also possible to use neural networks in unsupervised learning problems like reinforcement learning. By performing gradient descent on the weights using the model error, it is possible to tune $W$ until the model accurately approximates the original function between input and output vectors. A key observation here is that we expect this function to exist and be sufficiently represented in the training data. 

However, it took more than 60 years before increased in computational power and theoretical understanding finally made it possible to train large artificial neural networks (ANNs) on anything but trivial problems.

Replace the text \cite{example}, Figure~\ref{fig:picture},
Figures~\ref{fig:pictures}(a--b) and Table~\ref{tab:table1}.
\fi

\section{Deep Neural Networks}
The mathematical framework developed around biology-inspired Artificial Neural Networks dates back more than half a century \cite{mcculloch, rosenblatt}. However, only much more recent advancements in theoretical understanding and computational power have made it possible to successfully train large networks on complex tasks. Major theoretical milestones include the introduction of activation functions with non vanishing gradients \cite{relu}, layer regularization \cite{dropout} and advanced optimization algorithms \cite{adam}, while the computational side mostly benefited from the development of modern GPUs supporting highly parallel scientific computing at reasonable cost. The basic network structure in this study is a fully connected feed forward multi layer perceptron. All networks use three or more hidden layers, which classifies them as \emph{Deep Neural Networks}, according to conventional literature \cite{sergios}. While multiple hidden layers are not strictly necessary according to the Universal Approximation Theorem \cite{approximationtheorem}, they turn out to be very useful for training when combined with the aforementioned theoretical advancements. The basic structure of all networks is shown in figure \ref{nn1}. Hidden layers use rectified linear units. The output layer makes use of the softmax function to generate a probability like normalized output for different classes. At the most basic level, we simply use the space coordinates of four hits from the detector as inputs. For the two dimensional case study, that means each input is an 8-dimensional vector mapped onto a two dimensional normalized output representing good seeds and bad seeds. We use categorical cross entropy as a loss function and adaptive moment estimation is for optimization of the network weights.

\def\layersep{2.0cm}
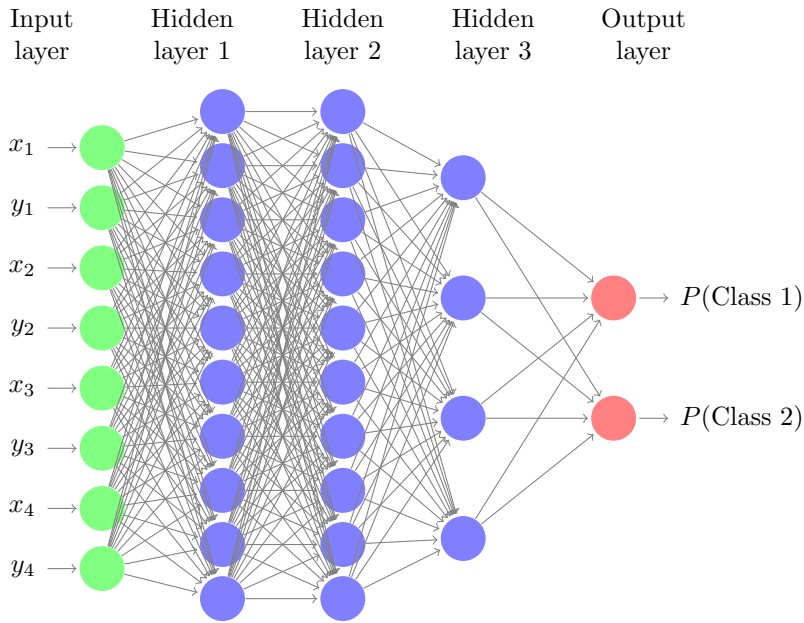
\begin{figure}[H]
\center
\begin{tikzpicture}[shorten >=1pt,->,draw=black!50, node distance=\layersep, scale=0.8]
    \tikzstyle{every pin edge}=[<-,shorten <=1pt]
    \tikzstyle{neuron}=[circle,fill=black!25,minimum size=17pt,inner sep=0pt]
    \tikzstyle{input neuron}=[neuron, fill=green!50];
    \tikzstyle{output neuron}=[neuron, fill=red!50];
    \tikzstyle{hidden1 neuron}=[neuron, fill=blue!50];
    \tikzstyle{hidden2 neuron}=[neuron, fill=blue!50];
    \tikzstyle{hidden3 neuron}=[neuron, fill=blue!50];
    \tikzstyle{annot} = [text width=4em, text centered]

    
    \foreach \id \name / \y in {$x_1$/1/1, $y_1$/2/2, $x_2$/3/3, $y_2$/4/4, $x_3$/5/5, $y_3$/6/6, $x_4$/7/7, $y_4$/8/8}
        \node[input neuron, pin=left:\id] (I-\name) at (0,-\y) {};

    \foreach \name / \y in {1,...,10}
        \path[yshift=0.5cm]
            node[hidden1 neuron] (H-\name) at (\layersep,-9/10*\y cm) {};
            
    \foreach \name / \y in {1,...,10}
        \path[yshift=0.5cm]
            node[hidden2 neuron] (J-\name) at (2*\layersep,-9/10*\y cm) {};
            
    \foreach \name / \y in {1,...,4}
        \path[yshift=0.5cm]
            node[hidden3 neuron] (K-\name) at (3*\layersep, - 2*\y cm) {};
            
    \node[output neuron,pin={[pin edge={->}]right:$P$(Class 1)}, right of=K-2] (O1) {};
    \node[output neuron,pin={[pin edge={->}]right:$P$(Class 2)}, right of=K-3] (O2) {};

    \foreach \source in {1,...,8}
        \foreach \dest in {1,...,10}
            \path (I-\source) edge (H-\dest);
            
       \foreach \source in {1,...,10}
        \foreach \dest in {1,...,10}
            \path (H-\source) edge (J-\dest);

       \foreach \source in {1,...,10}
        \foreach \dest in {1,...,4}
            \path (J-\source) edge (K-\dest);
                   
    \foreach \source in {1,...,4}
        \path (K-\source) edge (O1);

    \foreach \source in {1,...,4}
        \path (K-\source) edge (O2);

    \node[annot,above of=J-1, node distance=1cm] (hl2) {Hidden layer 2};
    \node[annot,left of=hl2] (hl1) {Hidden layer 1};
    \node[annot,right of=hl2] (hl3) {Hidden layer 3};
    \node[annot,left of=hl1] {Input layer};
    \node[annot,right of=hl3] {Output layer};
\end{tikzpicture}
\caption{A fully connected deep neural network with 8 input nodes, three hidden layers and two output nodes, as used for the two-dimensional case studies. Hidden layers in this figure are exemplary; each one contains hundreds of nodes in the actual model. For the three dimensional task, the input vector receives additional nodes. For classification of seeds in more than two groups, the output vector receives additional nodes.}
\label{nn1}
\end{figure}

\section{Case Studies}
While it was possible to jump immediately into particle and detector simulations, we decided to follow a more academic approach and start with simplified problems. Since we had no information regarding the complexity of this problem as a deep learning task, this approach gave us valuable estimates on the required model size and the necessary amount of training data. From there we continuously went on towards more and more realistic scenarios, eventually ending up in the production level simulation environment of ATLAS.
\subsection{Toy Environments}
To keep the scope of the environment and the model more manageable, we started with a seemingly much simpler task that is closely related to the heuristics of current detector software. We try to train a network that is capable of identifying whether or not four points are on a circle. This problem can easily be solved by hand and basically corresponds to solving the circle equation. To generate the training data, we used a simple program that generates circles of random radii at random center coordinates in a two-dimensional plane. Four points are then sampled on each circle, representing a ``good" seed. To generate ``bad" seeds, the fourth point is shifted away from the circle. The total training set contains 100k seeds of each class and is split into 75\% and 25\% sets for training and validation. With just a few minutes of training on a household GPU, the model can achieve accuracies beyond 95\%. This accuracy doesn't change even when we introduce some Gaussian smearing of the hit coordinates, to account for multiple scattering and detector inaccuracies, suggesting that the neural network is not actually solving the circle equation but actually working on some approximation. In a real particle physics detector, the interesting tracks are usually of very high momentum, so it is actually very rare to find tracks completing full circles inside the detector. So, in an attempt to move closer to what would be seen in a real experiment, we limit the sampling of points to one half of the circle. This small adjustment in the training data already leads to a significant increase in accuracy, with the model now identifying more than 99\% of the seeds correctly. We consider the toy model solved and now decide to move everything into three dimensions. The generation of training data remains the same, with the only difference being that circles are now generated with random orientation in three dimensional space. The model now requires 12 input nodes, leaving everything else unchanged. The introduction of the third dimension immediately pushes the maximum accuracy below 90\%, affirming that this is a substantial step in problem complexity for our deep learning model. Nonetheless the problem seems to be solvable by conventional approaches in deep learning.

\subsection{TrackML Detector Environment}
\label{trackmlsection}
Encouraged by the toy model results, we moved to a more sophisticated environment. The most readily available dataset stemmed from the Kaggle Machine Learning Challenge (TrackML) \cite{trackml}. The dataset contains simulated events from a hypothetical high energy particle physics experiment. Most importantly, since the data set was already geared towards machine learning, it contains not just the coordinates of hits in the detector, but also ground truth data for complete particle tracks. The task then simplifies towards selecting reasonable seeds from this dataset. Track reconstruction can start with seeds from the inside, closest to the beam-line, or from the outside. We start with the inside-out method in mind and select the first four hits from each ground truth track to generate good seeds. Bad seeds are generated by randomly switching out hits on a track seed with hits from other tracks. 

\begin{figure}[!htb]
  \centering
  \includegraphics[width=\linewidth]{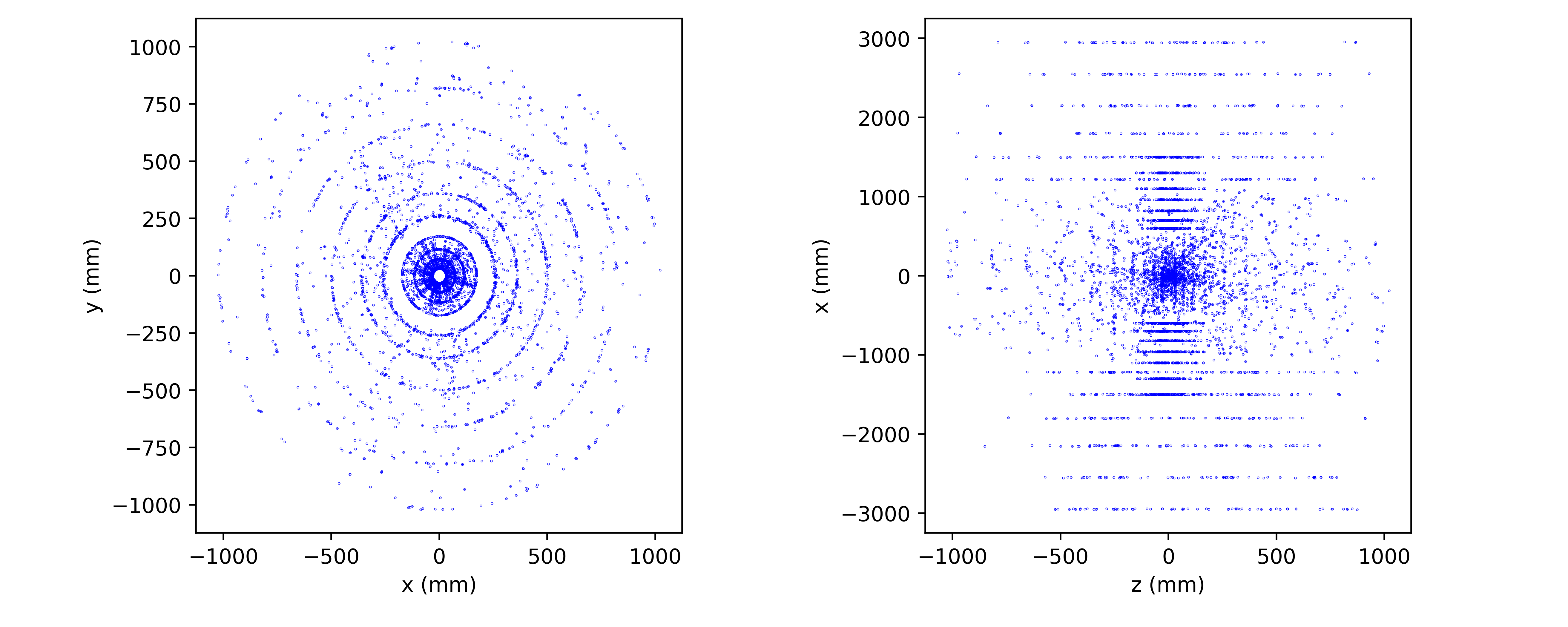}
  \caption{Detector hits from 500 simulated particle tracks. The raw data was generated using the same tools used for the TrackML challenge. The underlying detector geometry can be seen; it is already similar to what one would find in actual experiments such as ATLAS. The z-axis corresponds to the beam axis.}
  \label{fig:picture}
\end{figure}

The new dataset highlights that actual high energy tracks in a realistic particle detector are considerably simpler than what we saw in the toy models. We can now achieve up to 99.5\% accuracy on the validation set with just a few minutes of training. This is not unexpected. Tracks useful for physics usually come with high transversal momentum, making most tracks more similar to lines than a helix. We already saw that restricting spacial points to a half-circle greatly improves the accuracy of our model, so this additional reduction to an even smaller part of the circle is very likely to improve model performance. We also find that training on seeds generated from random hits along the entire track generates better results than training on seeds restricted to the first hits in the innermost layers. Even when the validation data is limited to the latter, the model trained on the ``larger" tracks performs better.

\begin{figure}[!htb]
  \centering
  \subfloat[]{\includegraphics[trim={0cm 0cm 0cm 0cm},  clip, width=0.45\linewidth]{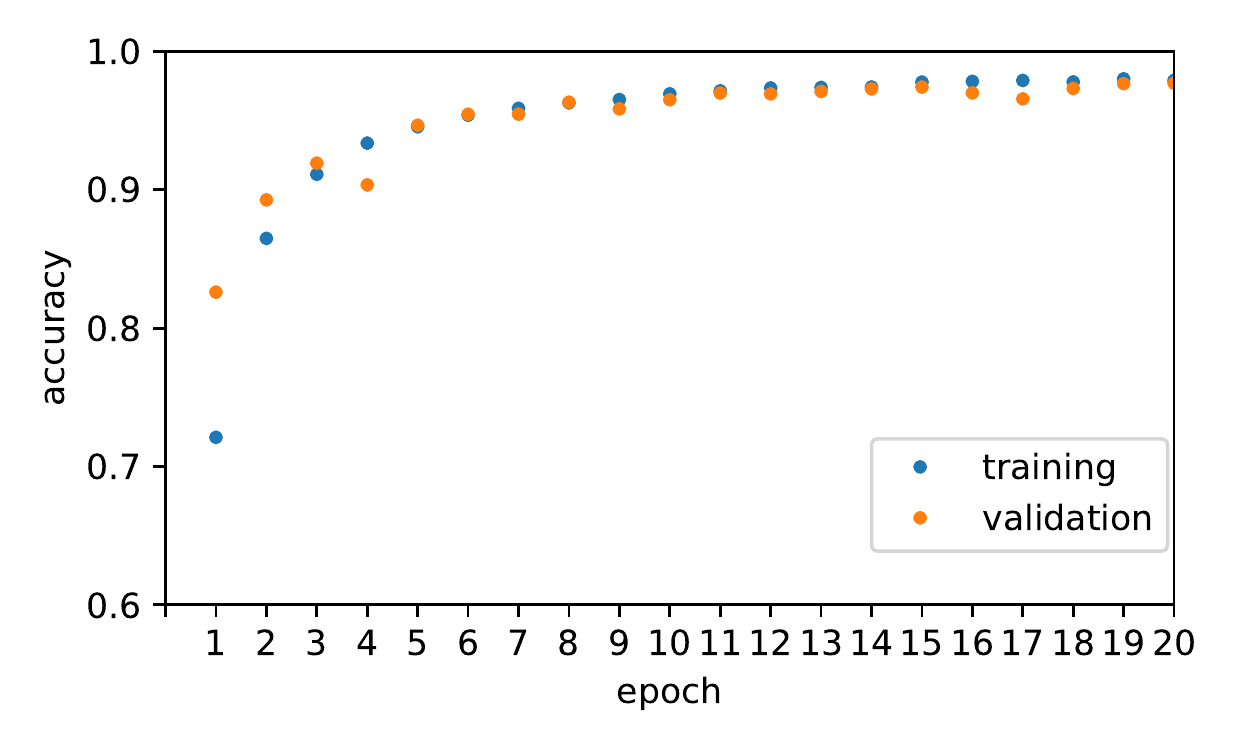}}
  \qquad
  \subfloat[]{\includegraphics[trim={0cm 0cm 0cm 0cm},  clip, width=0.45\linewidth]{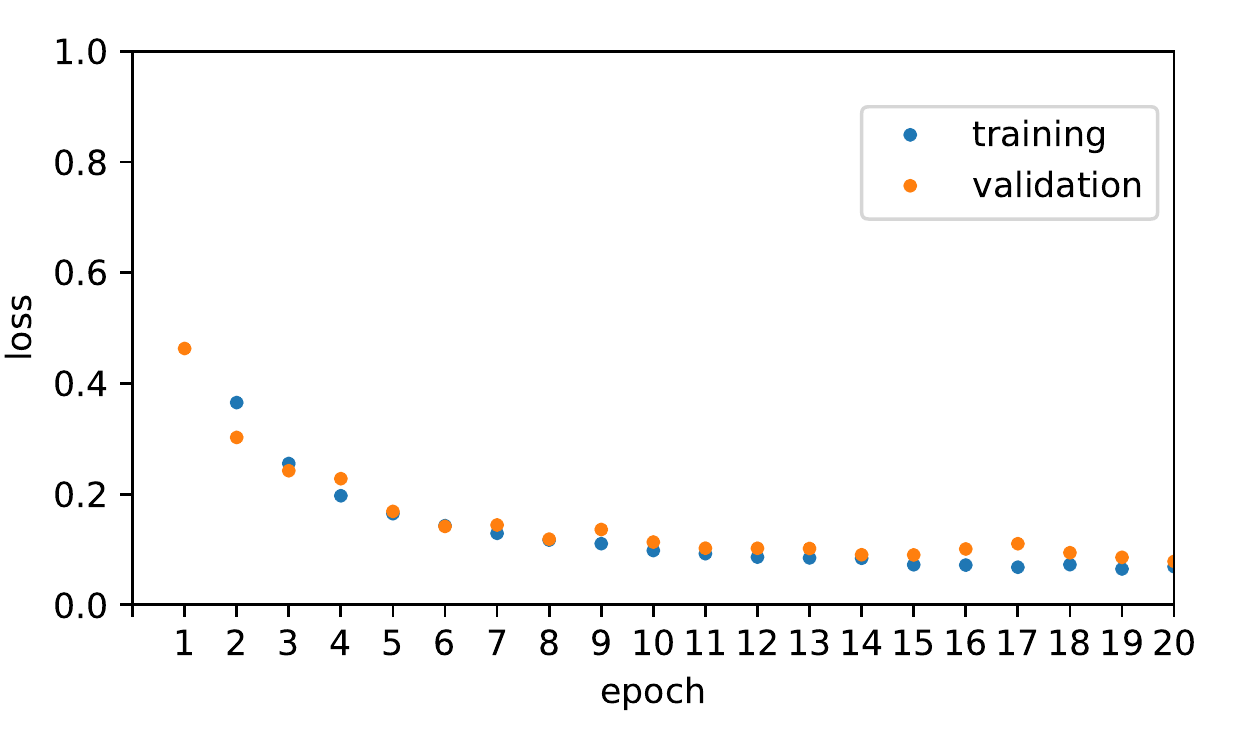}}
  \caption{Training history over 20 epochs showing the model prediction accuracy on the training and validation set. Note how both the loss and accuracy maintain a non-zero gradient all the way until the end of training while training and validation data stay close together. This suggests that we are not over-fitting yet and more epochs could result in even higher accuracy.}
  \label{fig:matrix}
\end{figure}

\subsection{ATLAS-style Detector Environment}
In the final step we actually move into the ATHENA track simulation environment of ATLAS. For training data generation, we gather 1020 particle tracks from 10 $t\bar{t}$-events and extract the spacial coordinates of every detector hit. To account for inevitable ambiguities in certain situations when deciding whether hits are on the same track, we also introduce another output category, ``medium", to account for these ambiguities. In this category, only one of the four hits per seed is wrong. Seeds in the \emph{bad} category now have at least two hits from other tracks. By only discarding bad seeds, we find that less than 0.5\% of actual on-track seeds would be eliminated by the model. So far, we randomly shuffled hits to generate bad seeds. But another important question is how the training process reacts to seeds that are more difficult to distinguish. To understand this, we create another training data set, but this time we take on-track hits and shift them slightly by a random amount to generate false seeds. Unsurprisingly, when the shift is too small the training fails and the model can no longer discern seed classes reasonably. In our dataset, it seems that shifts of a few dozen centimetres are necessary to achieve training success. The model trained on this dataset achieves slightly worse overall performance on the validation data set, but it is more capable when compared to the old validation set. This suggests that training on more difficult seeds can actually improve the overall model performance. The final model accuracy on both data sets is shown in table \ref{confmatrix1}. Each class in both validation sets contains roughly 25.000 seeds.

\begin{table}[!htb]
 \setlength{\extrarowheight}{5pt}
\begin{tabular}{cc}
\begin{tabular}{cc|c|c|c|@{}}
\multicolumn{1}{c}{} &\multicolumn{1}{c}{} &\multicolumn{3}{c}{Predicted class} \\ 
\multicolumn{1}{c}{} & 
\multicolumn{1}{c}{} & 
\multicolumn{1}{c}{Good} & 
\multicolumn{1}{c}{Medium} &
\multicolumn{1}{c}{Bad}  \\ 
\hhline{~~|-|-|-|}
\multirow[c]{2}{*}{\rotatebox[origin=tr]{90}{Actual class}}
& Good  & \cellcolor{blue!25}98.5\% & 1.5\%  & 0.1\%   \\
\hhline{~~|-|-|-|}
& Medium  & 3.5\% & \cellcolor{blue!25}95.7\%  & 0.8\%  \\
\hhline{~~|-|-|-|}
& Bad  & 0.2\%   & 3.2\%  & \cellcolor{blue!25}96.7\%  \\ 
\hhline{~~|-|-|-|}
\end{tabular}
& 
\begin{tabular}{cc|c|c|c|@{}}
\multicolumn{1}{c}{} &\multicolumn{1}{c}{} &\multicolumn{3}{c}{Predicted class} \\ 
\multicolumn{1}{c}{} & 
\multicolumn{1}{c}{} & 
\multicolumn{1}{c}{Good} & 
\multicolumn{1}{c}{Medium} &
\multicolumn{1}{c}{Bad}  \\ 
\hhline{~~|-|-|-|}
\multirow[c]{2}{*}{\rotatebox[origin=tr]{90}{Actual class}}
& Good  & \cellcolor{blue!25}98.5\% & 1.5\%  & 0.1\%   \\
\hhline{~~|-|-|-|}
& Medium  & 3.5\% & \cellcolor{blue!25}95.7\%  & 0.8\%  \\
\hhline{~~|-|-|-|}
& Bad  & 0.2\%   & 3.2\%  & \cellcolor{blue!25}96.7\%  \\ 
\hhline{~~|-|-|-|}
\end{tabular}
\\
\textbf{``Normal" training data} & \textbf{``Difficult" training data}
\end{tabular}
\captionsetup{width=\linewidth}
\caption{Confusion matrix for model performance on validation set. Ground truth seed class is labelled as actual class, model output as predicted class.}
\label{confmatrix1}
\end{table}

\subsection{Additional findings}
We find that additional layers can greatly speed up training. Our model usually makes use of three hidden layers as shown in figure \ref{nn1}. With an additional layer of the same size as the first two, the model can achieve the same accuracy with considerably less training. The final accuracy, however, is not much higher when compared to the smaller model. 
\begin{figure}[!htb]
  \centering
  \includegraphics[width=0.5\linewidth]{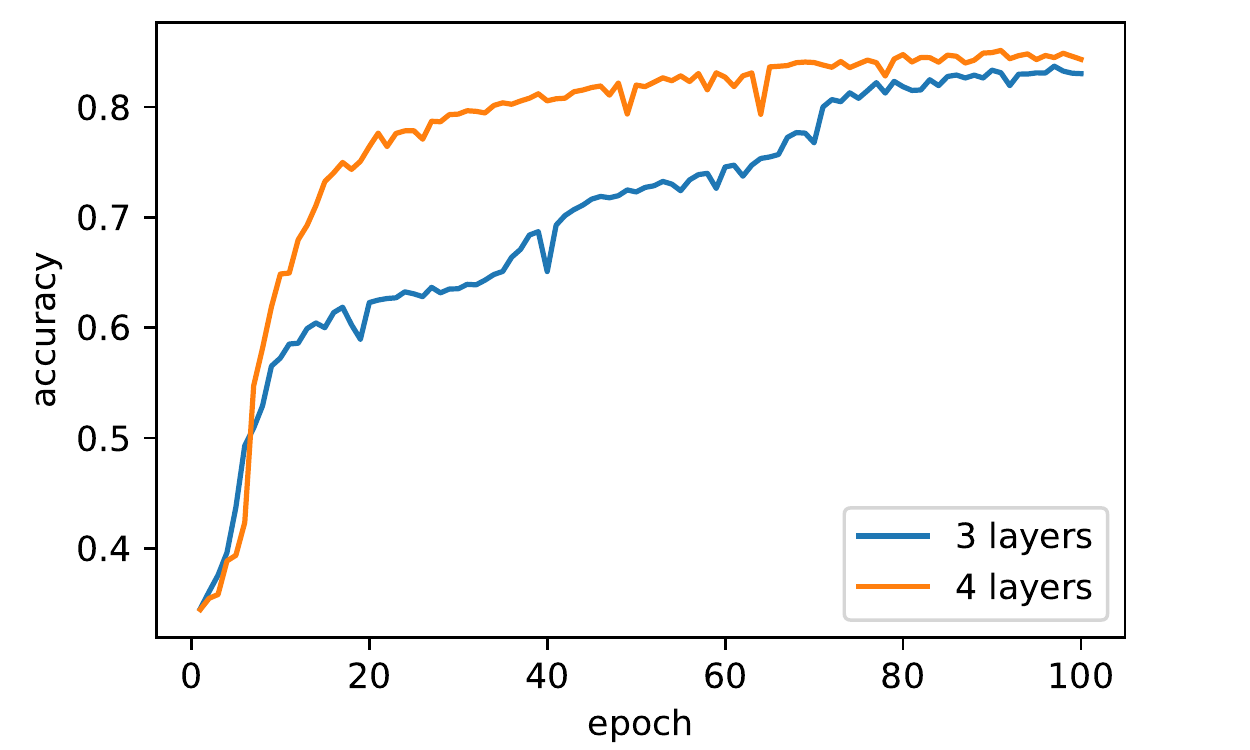}
  \captionsetup{width=0.5\linewidth}
  \caption{Accuracy of the DNN during training over 100 epochs.  An additional hidden layer clearly boosts the training speed, but the final accuracy benefits only by a small amount.}
  \label{fig:picture}
\end{figure}
We also find that a training set should contain at least $\mathcal{O}$(100k) seeds. A set of $\mathcal{O}$(10k) seeds turns out to be insufficient for training. In addition to these basic observations, we also find profound possibilities for future work: So far, we have seen that our  model can distinguish seed classes with considerable accuracy. However, due to the nature of deep neural networks, it is very hard to tell what the model is actually doing internally. Until now, we only know that it can learn to approximate some discriminatory hypersurface in a high-dimensional space. But we would also like to know how this abstract representation could represent the actual physics of particle tracks. To this end, we manually create seeds by setting three points on the xz-plane of the detector. Then we run the model on every possible position of the fourth point in the same plane and look at the probability to find a good seed, given that particular fourth hit-position. The result is shown in figure \ref{fig:matrix}. 

\begin{figure}[!htb]
  \centering
  \subfloat[]{\includegraphics[trim={2cm 0cm 0cm 1cm},  clip, width=0.4\linewidth]{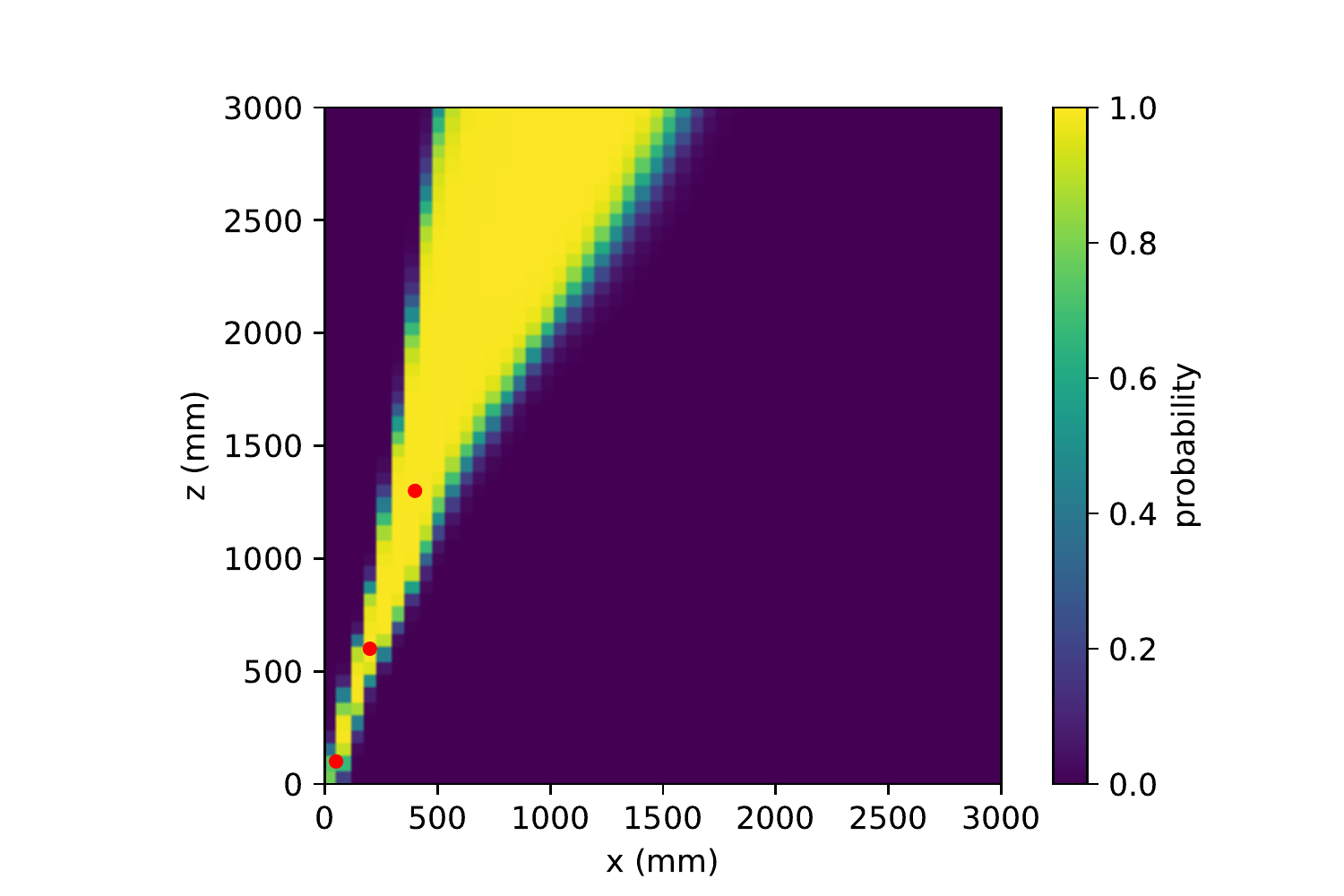}}
  \qquad
  \subfloat[]{\includegraphics[trim={2cm 0cm 0cm 1cm},  clip, width=0.4\linewidth]{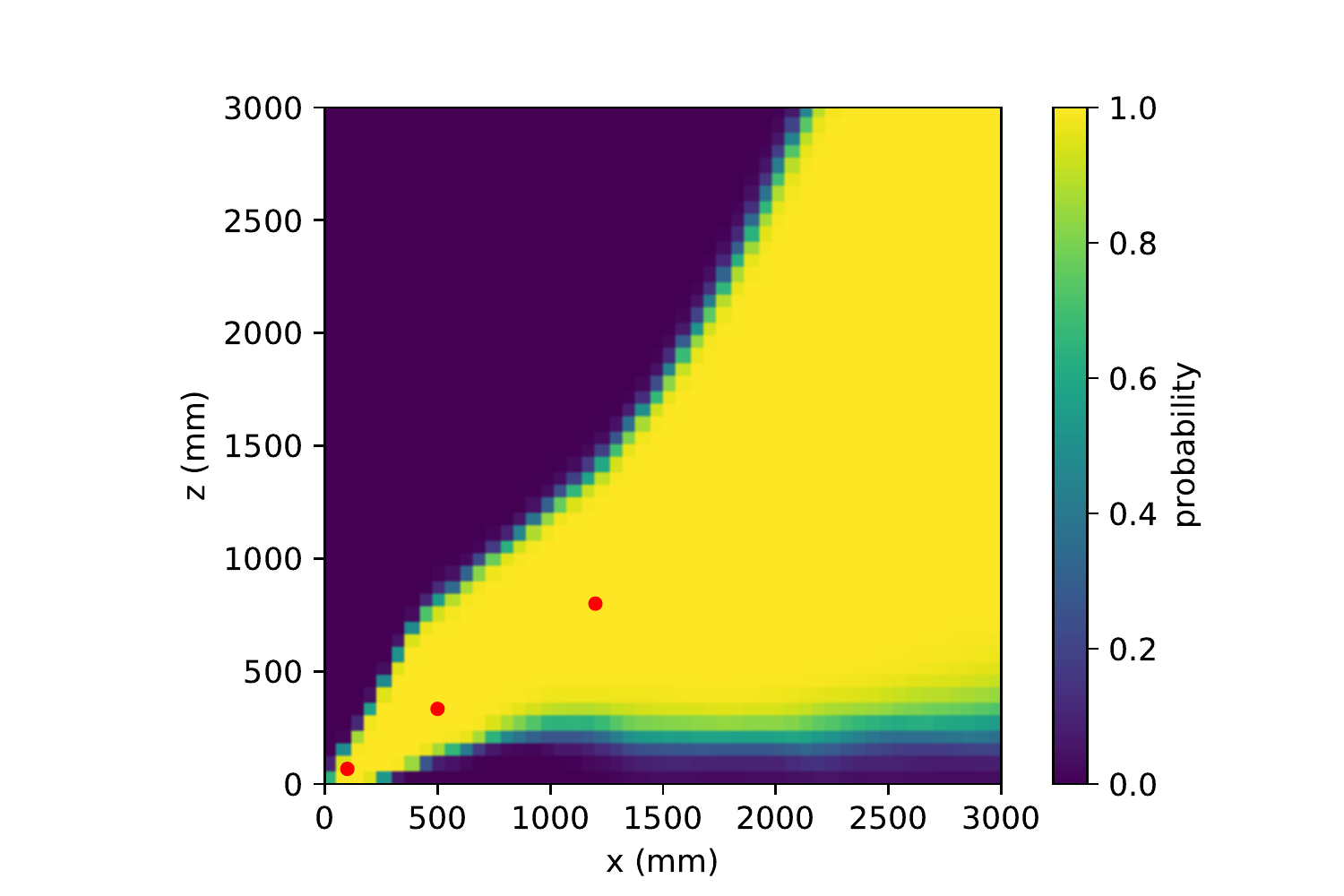}}
  \caption{Two examples of where the model estimates a probable fourth hit when given three hits of a seed. The positions of the manually placed hits are shown in red. Note that for lower transverse momentum tracks, the uncertainty in the model prediction grows considerably.}
  \label{fig:matrix}
\end{figure}

The model has obviously not just learned to classify seeds based on coordinates, but it has developed an internal approximation of where the entire track might be found. This has profound consequences for future work, as the same model might also be useful in the very next stage of track reconstruction. There it might act as a sort of Kalman-filter to find additional hits from an initial seed.


\section{Conclusions}

This work demonstrates that a deep neural network could be a powerful, easy to develop tool for track reconstruction. The only major requirement is a ground truth dataset of sufficient size. We show where major complications and benefits for training reside. Finally we present an analysis of the internal workings of the model and show that it might be useful not just for track seeding. We also haven't dived deeply into feature engineering yet. A real detector, for example, will provide more than just plain spacial coordinates for the hits. We expect that there is still plenty of room for improvement.


\Acknowledgements
I am grateful to Andreas Salzburger and Emmerich Kneringer for enabling and supporting this work. I also want to thank Noemi Calace for helping out with some ATLAS code. Furthermore, I want to thank the ATLAS group at the University of Innsbruck for their support.



\end{document}

%% file: econfmacros.tex
\def\beq{\begin{equation}}
\def\eeq#1{\label{#1}\end{equation}}
\def\eeqn{\end{equation}}

\def\beqa{\begin{eqnarray}}
\def\eeqa#1{\label{#1}\end{eqnarray}}
\def\eeqan{\end{eqnarray}}

\let\bar=\overbar

\def\Dslash{\not{\hbox{\kern-4pt $D$}}}
\def\dslash{\not{\hbox{\kern-2pt $\del$}}}

\def\msb{{\bar{\ssstyle M \kern -1pt S}}}

